\documentclass[aps,prb,reprint,superscriptaddress,raggedbottom]{revtex4-2}

\newcommand{\thetitle}{
Ultrafast gap dynamics upon photodoping the Mott-insulating phase \\ of a two-dimensional organic charge-transfer salt
}

\usepackage{graphicx}
\usepackage{amsmath}
\usepackage{amssymb}
\usepackage{xspace}
\usepackage{xcolor} 
\usepackage{upgreek}

\usepackage{times}
\usepackage{mathptmx}

\usepackage{titlesec}
\titleformat{\section}{\raggedright\bfseries}
{\arabic{section}.}{1em}{}
\titlespacing{\section}
{0pt}{6pt}{3pt}
\titleformat{\subsection}{\raggedright\itshape}
{\arabic{section}.\arabic{subsection}}{1em}{}
\titlespacing{\subsection}
{6pt}{6pt}{3pt}

\usepackage{soul} 
\soulregister\ref7
\soulregister\cite7
\soulregister\gls7
\soulregister\etal7

\usepackage[utf8]{inputenc}

\newcommand{\beq}{\begin{equation}}
\newcommand{\eeq}{\end{equation}}

\usepackage{pgffor} 

\newcommand{\usr}[2]{\ensuremath{#1_{\mathrm{#2}}}\xspace} %

\newcommand{\kETX}[1]
{$\kappa$-(BEDT-TTF)$_2$Cu[N(CN)$_2$]#1\xspace}

\newcommand{\kClfull}{\kETX{Cl}}
\newcommand{\kCl}{$\kappa$-Cl\xspace}
\newcommand{\kX}{$\kappa$-X\xspace}

\DeclareRobustCommand{\unitx}[1]{%
\ensuremath{\, 
\foreach \x/\y in {#1} {{\,\mathrm{\x}^{\y}}}}
\xspace
}
\newcommand{\unit}[1]{\ensuremath{\,{\mathrm{#1}}}\xspace}
\newcommand{\pcm}{\unitx{cm/-1}}
\newcommand{\pcmc}{\unitx{cm/-3}}
\newcommand{\pOhmcm}{\unitx{\Omega/-1,cm/-1}}
\newcommand{\mum}{\unit{\mu m}}
\newcommand{\muJcm}{\unitx{\mu J/,cm/-2}}

\newcommand{\fsec}{\unit{fs}}
\newcommand{\psec}{\unit{ps}}

\newcommand{\eVolt}{\unit{eV}}
\newcommand{\meVolt}{\unit{meV}}

\newcommand{\Kel}{\ensuremath{\,\mathrm{K}}}

\newcommand{\tisa}{Ti:Al$_2$O$_3$\xspace}

\newcommand{\simx}{\,{\sim}\,}

\newcommand{\dRr}{\ensuremath{\Delta R/R}\xspace}
\newcommand{\dRrel}{\usr{\Delta R}{rel}}
\newcommand{\Iref}{\usr{I}{ref}}

\newcommand{\wex}{\usr{\omega}{ex}}
\newcommand{\Fex}{\usr{F}{ex}}
\newcommand{\alphaex}{\usr{\alpha}{ex}}
\newcommand{\Nex}{\usr{N}{ex}}
\newcommand{\Nexm}{\usr{N'}{ex}}
\newcommand{\Ndim}{\usr{N}{d}}
\newcommand{\wz}{\ensuremath{\omega_{0}}\xspace}
\newcommand{\sigzn}{\ensuremath{\sigma_{0n}}\xspace}
\newcommand{\wzn}{\ensuremath{\omega_{0n}}\xspace}
\newcommand{\Gamn}{\ensuremath{\Gamma_{n}}\xspace}

\newcommand{\sigzX}[1]{\ensuremath{\sigma_{0,\mathrm{#1}}}\xspace}
\newcommand{\wzX}[1]{\ensuremath{\omega_{0,\mathrm{#1}}}\xspace}
\newcommand{\GamX}[1]{\ensuremath{\Gamma_{\mathrm{#1}}}\xspace}

\newcommand{\epsr}{\ensuremath{\hat \varepsilon}\xspace}
\newcommand{\epsri}{\ensuremath{\hat \varepsilon_{\infty}}\xspace}

\newcommand{\ffm}{Physikalisches Institut, J. W. Goethe-Universität, D-60438 Frankfurt am Main, Germany}

\newcommand{\stutt}{
1. Physikalisches Institut, Universität Stuttgart, D-70550 Stuttgart, Germany}
\begin{document}

\title{\thetitle}
\author{Konstantin Warawa}
\affiliation{\ffm}
\author{Yassine Agarmani}
\affiliation{\ffm}
\author{Harald Schubert}
\affiliation{\ffm}
\author{Martin Dressel}
\affiliation{\stutt}
\author{Michael Lang}
\affiliation{\ffm}
\author{Hartmut G. Roskos}
\affiliation{\ffm}
\author{Mark D. Thomson}
\affiliation{\ffm}

\begin{abstract}
We investigate experimentally the ultrafast changes in the spectral response of the Mott insulator \kClfull (\kCl) upon photodoping with intense excitation at 1.6\eVolt and probing with continuum pulses simultaneously covering both the terahertz and infrared (IR) ranges (from 0 to $0.6\eVolt$).  
A quantitative analysis of the differential reflectivity using a multi-band Lorentzian model provide absolute changes in spectral weights and objective global time constants for the relaxation vs. temperature.
The transient conductivity spectra deduced from the analysis 
suggest that the transient photoinduced spectral weight is dominated by a progressive closure of the Mott gap with increasing excitation density, i.e. due to changes in the inter-Hubbard-band absorption by the remaining singly occupied states. 
We critically examine this scenario compared to that proposed previously, whereby the low-energy spectral weight is attributed to a Drude-like response of photoexcited doublons/holons.
We also consider the observed slowing down of the  relaxation rate with increasing excitation density, and temperature dependence of the initial doublon/holon density in terms of the phonon-mediated gap recombination model.
\end{abstract}

\maketitle

\section{Introduction}

The physics of quantum materials with strong electron correlations, such as low-dimensional Mott-Hubbard systems \cite{dressel04,toyota07}, continues to be highly active field of research, and while much progress has been made in the understanding of the ground-state (GS) properties, the non-equilibrium response due to intense, ultrafast excitation (either with strong, sub-gap electric fields \cite{yamakawa17,kawakami18,yamakawa21}, nonlinear phonon excitation \cite{buzzi20}, or photodoping via interband excitation \cite{kawakami09,toda11,tsuchiya_threepulse21}) remains an exciting and challenging frontier. %
An exceptional aspect of photodoping for strongly correlated systems is the possibility to reach non-thermal/hidden states \cite{stojchevska14}, as a change in the state occupation can lead to a drastic, concerted renormalization of the electronic energy-state distribution \cite{eckstein11,eckstein13,murakami_arxiv23} (and hence also exhibit profound effects due to coupling with the lattice/spins \cite{grandi21}), in contrast to conventional semiconductors where band renormalization effects are often minor.
Moreover, photodoping serves as a complementary approach to conventional charge doping (e.g. chemical or electrostatic \cite{kawasugi11}) for studying Mott systems, in that one preserves the total charge concentration (e.g. half-filling).

Here we investigate the two-dimensional charge transfer salts
\kETX{X} (\kX), which are Mott systems that are highly susceptible to (chemical) pressure and doping, 
and hence possess a rich phase diagram (including superconducting and AF magnetically ordered phases, and a Mott metal-insulator transition with a critical endpoint) %
\cite{lefebvre2000a,dressel04,toyota07,kanoda11,basov11,dressel20}. 
We focus on the ultrafast response to photodoping in the Mott insulator \kCl for temperatures $T<100\Kel$.
The excited-state (ES) dynamics in \kCl have been intensively studied previously, using either 
optical-interband 
\cite{kawakami08,naito08,kawakami09,toda11,tsuchiya19,tsuchiya_threepulse21} 
or strong-field 
\cite{yamakawa17,kawakami18,yamakawa21} 
excitation, with the response probed either via the low-energy excitations across the mid-infrared (IR) \cite{kawakami08,kawakami09,yamakawa17,kawakami18} or changes in higher-energy interband absorption \cite{naito08,toda11,tsuchiya19,tsuchiya_threepulse21,yamakawa21}.
The studies covering the mid-IR range observed signatures for the bleach of the GS transition from the lower- to upper-Hubbard bands (LHB, UHB), and a photoinduced spectral weight at lower energy which was attributed to the mid-IR tail of a Drude-like response from photoexcited carriers (doublon/holons). Moreover, it was speculated that this is accompanied by a collapse of the Mott gap \cite{yamakawa17}, although no excitation dependence or threshold for such a collapse was elucidated.
Here we build on these studies, by extending the probe spectral range to simultaneously cover both the terahertz (THz) and mid-IR regions with femtosecond (fs) continuum pulses and ultra-broadband spectral detection \cite{thomson23}.  Moreover, we perform a rigorous analysis of the differential reflectivity spectra to yield the underlying transient ES conductivity spectra, as a function of excitation fluence (density), probe delay and GS temperature.
These results show that while the photoinduced spectral weight indeed begins to rise in the mid-IR already at low fluence, it deviates strongly from a Drude-like frequency dependence, and rather suggests it is due to a photoinduced change in the interband absorption of remaining singly occupied states 
(``singlons''), with a gap that progressively closes with increasing fluence.  
This constitutes a significantly different scenario for ongoing theoretical treatments of the effect of photodoping in such narrow-gap Mott systems. %

While the observed temperature dependence of the relaxation rate is similar to that found in previous all-optical pump-probe studies %
\cite{naito08,toda11,tsuchiya19}, we find here that the relaxation slows down with increasing excitation density.  
Based on the established phonon-mediated gap recombination model \cite{kabanov99,kabanov05}, this may be due to an increase in the phonon bottleneck as the gap energy diminishes.
We also revisit the temperature dependence of the initial excitation density \cite{toda11} (following the rapid sub-picosecond processes where carrier multiplication and energy partitioning to high-energy phonons occurs), and show that the dependence is indeed quite consistent with the predictions of the gap-recombination model.


\section{Ground-state properties}

\begin{figure*}[ht]
\centering
\includegraphics
[width=\textwidth]
{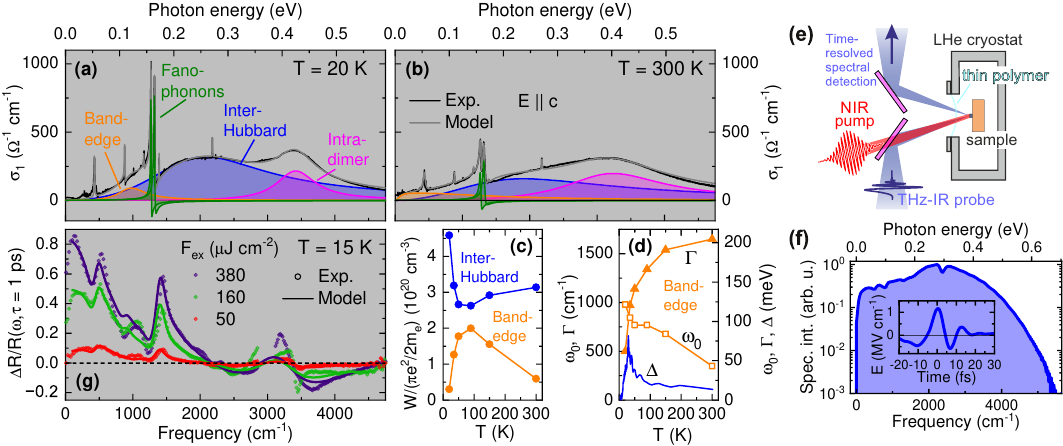}
\caption{
(a,b) Experimental ground-state conductivity spectra of \kCl and multi-band Lorentzian fits for (a) $T=20\Kel$ and (b) $T=300\Kel$, polarization parallel to the $c$-axis.
(Experimental data originally published in \cite{faltermeier07}; see Supplementary for full set of spectra vs $T$). Band models include Fano factors for selected phonons -- see text. %
(c,d) Selected band parameters vs. $T$: (c) spectral weight of inter-Hubbard and band-edge components (normalised to equivalent density assuming free electron mass); (d) Lorentzian peak frequency $\wz$ and bandwidth $\Gamma$ for the band-edge component.  
Also included is the nominal gap 
$\Delta=2\partial_\beta(\ln{\rho})$ %
from previously reported DC transport measurements \cite{tomic13,pinteric15}.
(e) Schematic of optical-pump THz-IR-probe experimental geometry. %
(f) Intensity spectrum of THz-IR probe pulses (inset shows reconstructed temporal electric field at focal plane \cite{thomson23}). %
(g: bottom left) Exemplary excited-state differential reflectivity spectra $\dRr$ (points) at $T=15\Kel$ and delay $\tau=1\psec$ after excitation 
($\hbar\wex=1.6\eVolt$) for the given values of excitation fluence $\Fex$, and fitted spectra (solid curves) using excited-state multi-band Lorentzian model.  Common frequency scale as in (a).
} \label{fig:gs}
\end{figure*}

The low-energy ground-state (GS) excitations in \kCl have been well characterized in previous reports \cite{kornelsen92,sasaki04,faltermeier07,dumm09,elsasser12,tomic13,vlasova09,vlasova2011}.
As shown in Fig.~\ref{fig:gs}(a), at low temperature the conductivity spectrum $\sigma(\omega)$ features two main, broad bands in the mid-IR, which have been assigned to (i) the inter-Hubbard-band transition 
(between the LHB and UHB, centered at $\wz= 2100\pcm$) %
and (ii) intra-dimer charge transfer ($\wz = 3400\pcm$) \cite{faltermeier07,ferber14}.  These bands can be fit well using a Lorentzian model for the conductivity $\sigma=\sigma_1+i\sigma_2$, i.e.
$\sigma_n(\omega)=i\Gamma_n\omega\sigma_{0n} (\omega_{0n}^2-\omega^2+i\Gamma_n\omega)^{-1}$.  
A series of phonons are also present, some of whose non-IR-active symmetry is lifted via intensity borrowing from electron-molecular-vibrational (EMV) coupling \cite{rice80,painelli86,dressel04}, in particular the strong doublet at $\omega_{0}=1270$ and $1320\pcm$.  Consistent with significant electron-phonon (e-ph) coupling, these phonons exhibit rather a Fano lineshape %
$\tilde\sigma_{n}(\omega)=(q_n+i)^2/(1+q_n^2)\sigma_{n}(\omega)$ ($1/q_n\neq 0$) \cite{fano61,kornelsen92,sedlmeier12,elsasser12}, due to coherent interference with the overlapping bands, as seen in Fig.~\ref{fig:gs}(a).
Even at low $T$, a remaining spectral weight contribution is present at lower frequencies, which we also fit with a Lorentzian band (also included in Fig.~\ref{fig:gs}(a)). As motivated below, we refer to it as a ``band-edge'' (BE) contribution, speculating that it augments the main inter-Hubbard band, accounting for the detailed energy-state distribution of the LHB/UHB.
These bands are retained with increasing $T$ (even up to $T=300\Kel$,  Fig.~\ref{fig:gs}(b)), although with a modification in their parameters (see Supplementary for spectra at additional $T$ values).  
One sees that the clear gap in $\sigma_1(\omega\rightarrow 0)$ at $T=20\Kel$ has been filled in, with a plateau of $\sigma_1\simx 50\pOhmcm$.
Typically, one would attribute the spectral weight in this region to a Drude-like contribution from thermally excited carriers. 
While the spectral range here is limited to $\omega\gtrsim 50\pcm$, the DC conductivity at $T=300\Kel$ is $\sigma_1(0)\simx 5\pOhmcm$ \cite{pinteric15} -- hence any free-carrier contribution would need to be of the unconventional, modified Drude type (such as Drude-Smith with a large scattering bias -- see Supplementary).  Here we instead choose to reproduce this spectral weight by red-shifting and broadening the BE band.

In Fig.~\ref{fig:gs}(c,d) we plot selected band parameters vs. $T$.
Fig.~\ref{fig:gs}(c) shows the spectral weights 
($W_n=\int \mathrm{d}\omega \, \sigma_{1n}(\omega)$ \cite{dresselbook}) for the main inter-Hubbard (H) and BE components, normalized to the effective density assuming the free electron mass (for comparison, note that the density of dimers in \kCl is $\usr{N}{d} = 6.2\cdot 10^{20}\pcmc$ \cite{williams90}). 
One sees that while $\usr{W}{H}$ drops significantly going from $T=15\Kel$ to $50\Kel$, this is matched by an increase in $\usr{W}{BE}$. While one could consider that the BE component is associated with band orbitals distinct from the HBs which become increasingly populated, this seems unlikely, given that the thermal energy $kT$ over some $10\Kel$ scale corresponds only to a few meV, while the H- and BE-band peaks are separated by $>100~\meVolt$.  Hence, it appears more likely that both bands pertain to orbital-bands with strong electronic correlations (which can inherently undergo energetic reconfiguration on scales larger than changes in $kT$), supporting our assertion that the BE band is associated with the HBs.  
In addition, Fig.~\ref{fig:gs}(d) shows that the BE band undergoes a significant red-shift (and broadening) with increasing $T$.  Within our current hypothesis, this then reflects changes in inter-Hubbard absorption due to a (partial) closure of the Mott gap.

In Fig.~\ref{fig:gs}(d), we also include the estimate of the nominal gap energy based on previously reported measurements of the DC resistivity $\rho(T)$ \cite{tomic13,pinteric15}, i.e. obtained from the expression 
$\Delta(T)=2\partial_\beta(\ln{\rho(T)})$ %
($\beta=(kT)^{-1}$).
While it is difficult to directly compare this gap estimate with the BE band parameters directly, one can perceive a certain degree of  correspondence (for $T>30\Kel$), with the implied gap decreasing in both cases with increasing $T$ (noting that both $\wz$ and $\Gamma$ should be considered for assessing the gap energy from the optical data).
The roll-off in the transport gap for $T<30\Kel$ is in a range where $\rho$ has already reached high relative values \cite{pinteric15}, where a small concentration of residual carriers (e.g. due to mid-gap impurity states) may influence the result.
In any case, one sees that for $T>100\Kel$, a value of $\Delta\simx 20\meVolt$ is predicted, which we will see is approximately consistent with the photodoping concentrations determined below.

\section{Excited-state dynamics}

\subsection{Experimental details and spectral analysis}

Details of the optical-pump THz-IR-probe setup (based on a 1-kHz \tisa femtosecond amplifier laser) are given in the Supplementary, with a scheme showing the sample region in Fig.~\ref{fig:gs}(e).  In brief, the probe pulses are generated in a two-color air plasma \cite{thomson10,thomson23}, covering the spectral range from $\simx0\mbox{-}0.6\eVolt$ (Fig.~\ref{fig:gs}(f)), with minimal dispersion in the beam path to obtain near-transform-limited pulses on the \kCl samples, which are mounted in a liquid-helium optical cryostat (equipped with a thin polypropylene window of thickness 15-30~\mum).  The samples have lateral dimensions of $\simx 500\mum$, and were selected to have smooth as-grown surfaces ($a$-$c$ plane). 
The pump pulses ($\hbar\wex=1.6\eVolt$, $\simx 150$-fs duration) are introduced collinearly with the probe using an indium-tin-oxide beam-combiner before the sample, arriving a time $\tau$ before the probe pulse, with a beam size sufficiently large to have a uniform excitation fluence over the probe focal beam profile.  
Here we employed s-polarization for both pump/probe fields, parallel to the $c$-axis of the \kCl samples, with incidence angle $\theta = 18^{\circ}$.
Following reflection, the probe pulses are reimaged to a subsequent focal plane and measured using field-induced second-harmonic detection (FISH) \cite{thomson23} to yield the intensity spectrogram $I(\omega,t)$.
As the FISH gate pulse is sufficiently long (150~fs) compared to the THz-IR pulse, one can use a constant detection time $t=t_0$ and still capture the full probe spectrum $I(\omega,\tau; t_0)$, with a frequency resolution of $\delta\omega \simx 10\meVolt$.
An electromechanical shutter in the pump beam allows tracking of the reference reflected spectrum $\usr{I}{ref}(\omega,t_0)$ during measurements, and to calculate the relative differential reflectivity 
$\dRrel(\omega,\tau)\equiv \Delta R(\omega,\tau)/R(\omega)$ via 
$\dRrel=(I-\Iref)/\Iref$.

Example excited-state (ES) spectra $\dRrel(\omega,\tau=1\psec)$ are shown in Fig.~\ref{fig:gs}(g) for $T=15\Kel$ and different excitation fluences \Fex %
(the GS reflectivity spectrum for $T=20\Kel$ is given in the Supplementary).
A comparison of the features in the GS conductivity spectrum (Fig.~\ref{fig:gs}(a)) indicate a predominant bleach ($\dRrel<0$) in the mid-IR range (as seen in previous studies on \kX in the insulating phase, with manually tuned, narrow-band probe pulses 
\cite{kawakami09,yamakawa17}), and enhanced spectral weight ($\dRrel>0$) at lower frequencies (as well as derivative-like features about each strong phonon and near the peak of the intra-dimer band -- addressed below).

In order to reveal the quantitative conductivity changes underlying the \dRr spectra, we apply a model based on photoinduced changes of the GS band parameters -- i.e. $\sigzn\rightarrow \sigzn'(\tau)$, $\wzn\rightarrow \wzn'(\tau)$, $\Gamn\rightarrow \Gamn'(\tau)$ -- to yield excited-state components $\sigma'_n(\omega,\tau)$, as well as allowing for a change in the relative permittivity contribution from higher-lying resonances 
$\epsri \rightarrow \epsri'(\tau)$. 
Here we carefully selected a small set of band parameters to change from their GS values (see below), in order to avoid overfitting of the spectra while still capturing the main features. %
The ES conductivity model applies to the photoinduced changes directly at the sample surface. 
To calculate the model $\dRrel$, we first account for the excitation depth profile, i.e. taking
$\epsr'(\omega,z)=\epsr(\omega) + 
\Delta\epsr(\omega)\cdot e^{-\alphaex z}$, where 
$\epsr(\omega)=\epsri-i\sigma/\varepsilon_0\omega$ is the GS relative permittivity (likewise for the ES quantities $\epsr'$ and $\sigma'$) and $\Delta\epsr=\epsr'-\epsr$. %
The complex GS reflection coefficient $r(\omega)$ is calculated using Fresnel equations valid for complex $\epsr$ and oblique incidence. %
For the ES reflection coefficient $r'(\omega,\tau)$, one can 
derive an analytic expression taking the depth profile into account, even for oblique incidence in the case of s-polarized probe light (see Supplementary and \cite{khachatrian15}).  Here we take $\alphaex=3\mum$, based on an analysis of the thin-film transmission spectrum of \kCl (for $c$-axis polarization) given in the Supplementary of \cite{kawakami20}. %
%
Moreover, to account for the finite frequency resolution, we convolve $r'(\omega,\tau)$ with the FISH response function (a Gaussian with half-width $\delta\omega$).
One can show that applying this to the \textit{field} (as opposed to intensity) coefficients, 
$r(\omega)$ and $r'(\omega)$, accounts also for the effect of the finite reflective group delay \cite{Meng2015} for the time-gated TFISH spectrum, i.e. temporal shifts of frequency components away from $t=t_0$, which are not negligible, especially around strong resonances.  Finally one calculates the reflectivity $R'=|r'|^2$ and $\dRrel=(R'-R)/R$.

Example model fit spectra are also shown in Fig.~\ref{fig:gs}(g), where we allow only the following ES band parameter changes: (i) inter-Hubbard band strength in terms of the peak band conductivity $\sigzX{H}'$ (bleach); (ii) all parameters of the band-edge component ($\sigzX{BE}'$,$\wzX{BE}'$,$\GamX{BE}'$), and (iii) the bandwidth of the intra-dimer transition ($\GamX{D}'$).  The last parameter was included based on the observed GS $T$-dependence for the intra-dimer band (compare Fig.~\ref{fig:gs}(a,b)), and was found to assist reproducing the peak feature in \dRrel in the range $\omega=3000\mbox{-}3500\pcm$.
The model spectra are seen to reproduce the main features in the experimental \dRrel spectra across the full spectrum.  Note that the signatures near strong GS phonon bands (particularly pronounced for the phonons at 430, 880 and $1300\pcm$) are also present in the model curves, even though the phonon band parameters are kept constant at their GS values.  Instead, these signatures manifest due to the non-additive nature of the reflectivity with respect to overlapping conductivity components.
This emphasizes one important aspect of the foregoing spectral analysis, as a direct assessment of \dRrel in these spectral ranges could lead one to incorrectly assert that the phonons are strongly perturbed by the photodoping.

\begin{figure*}[t]
\centering
\includegraphics
[width=\linewidth]
{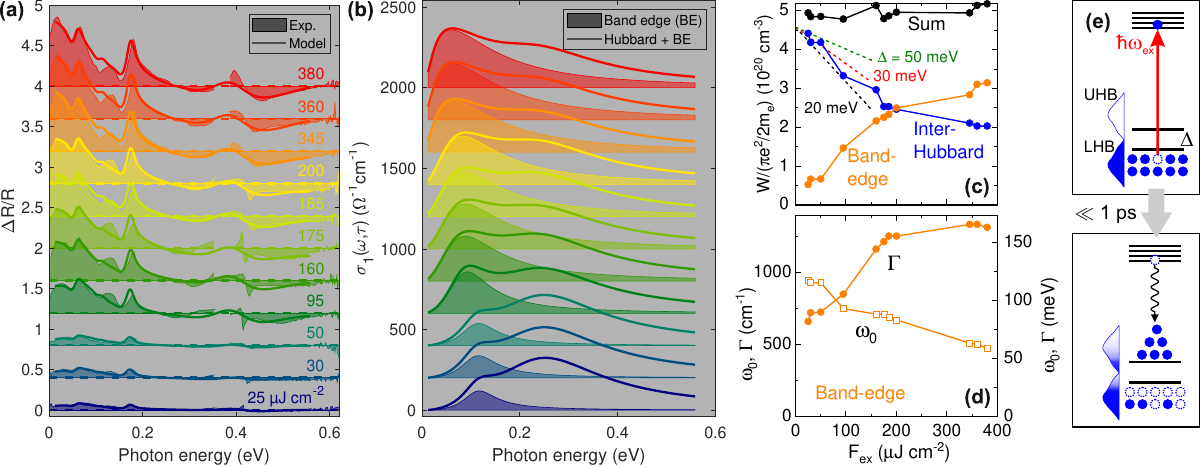}
\caption{
Excited-state spectra for \kCl vs. excitation fluence $\Fex$ for $T=15\Kel$ and delay $\tau=1\psec$ after excitation. (a) Differential reflectivity spectra $\dRr$ (\Fex labeled for each curve, vertically staggered for clarity), and model fits (as per Fig.~\ref{fig:gs}(e)). %
(b) Corresponding model conductivity spectra: Lorentzian components from band-edge alone (filled regions) and sum of inter-Hubbard and band-edge components (curves). %
(c,d) Corresponding excited-state Lorentzian model parameters: (c) selected spectral weights (and their sum) and (d) Lorentzian peak frequency $\wz'$ and damping $\Gamma'$ for band-edge component vs. \Fex (compare with ground-state parameters vs. $T$ in Fig.~\ref{fig:gs}(c,d)).
Included in (c) are predictions of the inter-Hubbard band spectral weight (dashed lines) assuming a simple two-level bleach 
$\usr{W'}{H}=\usr{W}{H}(1-2\Nexm/\Ndim)$, with a holon-doublon excitation density $\Nexm=(\hbar\wex/\Delta)\Nex$ i.e. following complete conversion of initial excitations at $\hbar\wex$ to those with the gap-edge energy $\Delta$. %
(e) Schematic of multiple doublon-holon generation following optical excitation, whereby each higher-energy electronic excitation quantum at $\hbar\wex$ rapidly distributes into multiple excitations at the gap edge. %
} \label{fig:fluence}
\end{figure*}

\subsection{Excitation density dependence and gap-closure model}

In Fig.~\ref{fig:fluence}, we investigate the excitation density dependence in more detail.  Fig.~\ref{fig:fluence}(a) shows a set of experimental and fitted spectra at $T=15\Kel$ for different fluences \Fex, at a delay $\tau=1\psec$, i.e. after initial fast electronic scattering processes have occurred, but where ES population relaxation is not yet significant (see below).  As per the GS (Fig.~\ref{fig:gs}(c,d)), we plot selected model ES Lorentzian band parameters in Fig.~\ref{fig:fluence}(c,d).  In (c), one sees the analysis
yields a significant drop in the spectral weight \usr{W'}{H} of the inter-Hubbard band, which begins close to a linear dependence on \Fex before the onset of saturation (reaching $\usr{W'}{H}/\usr{W}{H}=0.44$ at the highest fluence). One also observes that the spectral weight of the BE component increases in essentially quantitative proportion (as seen by the sum $\usr{W'}{H}+\usr{W'}{BE}$ shown). This is reminiscent of the trend seen in GS vs $T$ in Fig.~\ref{fig:gs}(c,d), although here we anticipate a much higher excitation density than for the thermal GS case (also discussed below).  Moreover, in Fig.~\ref{fig:fluence}(d), one sees that the BE undergoes a significant red shift and broadening with increasing \Fex.
However, the precise relative parameter changes deviate from those vs. $T$ in the GS, demonstrating that the ES is not simply described by heating effects.

Before interpreting these spectral changes further, we discuss the nature of the optical excitation and expected doublon/holon excitation densities vs. \Fex.
The initial electronic excitation at $\hbar\wex=1.6\eVolt$ is well above the inter-Hubbard energy, and should rather involve interband absorption from lower-lying bands, associated with HOMO[-$n$] orbitals of the organic (BEDT-TTF) molecules \cite{naito08,tsuchiya19} (although DFT calculations suggest the first LUMO above the Hubbard bands might also be reachable \cite{ferber14}).  Electron-electron scattering can rapidly allow for these excitations to redistribute their energy with additional electrons (akin to Auger processes), especially given the dense manifold of overlapping occupied bands below the GS Fermi energy \cite{ferber14}.
Hence, the prevailing hypothesis is that the initial density of excitations $\Nex$ with energy $\hbar\wex$ should break down to generate a multiplied population $\Nexm$ of excitations residing directly above the Mott gap with energy ${\gtrsim}\Delta$ \cite{yamakawa17,werner14}, as depicted schematically in Fig.~\ref{fig:fluence}(e). In the limit where losses from, e.g., inelastic phonon scattering are negligible, we then expect $\Nexm=(\hbar\wex/\Delta)\Nex$.
For our experiments, we have $\Nex=\alphaex\Fex/(\hbar\wex)$ reaching $\Nex=4.6\cdot 10^{18}\pcmc$ at the highest fluence $\Fex=380\muJcm$, i.e. exciting an initial fraction $\Nex/\Ndim=0.7\%$ of the dimer moieties (which will then be multiplied by up to a factor $\hbar\wex/\Delta \gg 10$). 
In Fig.~\ref{fig:fluence}(c), we also include the predicted spectral weight $\usr{W'}{H}(\Fex)$ of the inter-Hubbard band for the different possible values of the gap $\Delta$, assuming a simple bleach model where $\usr{W'}{H}=\usr{W}{H}(1-2\Nexm/\Ndim)$ (the factor 2 accounting as usual for both reduced absorption and stimulated emission).
As can be seen, the experimental bleach is consistent with a gap energy in the range $\Delta=20\mbox{-}30\meVolt$ (assuming the maximum multiplication factor), close to that  estimated from DC transport for $T>100\Kel$ (Fig.~\ref{fig:gs}(d)), whereby the ES electron temperature is indeed expected to be well above the GS temperature $T=15\Kel$.

Returning to the ES spectral changes associated with the BE component, one can more readily perceive these effects in Fig.~\ref{fig:fluence}(b), where we plot the components of the model ES conductivity spectra $\sigma'(\omega)$ for each \Fex, where we include both $\usr{\sigma'}{BE}$ alone and $\usr{\sigma'}{H}+\usr{\sigma'}{BE}$.  The progressive shift of the low-frequency edge is clearly evident.
The analysis employed here -- i.e. modifying only existing GS bands in the ES -- implicitly promotes the notion that the dominant spectral changes are associated with interband absorption for the remaining singlons in the LHB, which are affected by the ES environment, at least in terms of the increase in average electron temperature, assuming electronic thermalization across the LHB and UHB has taken place to a reasonable degree.
This clearly contrasts with the case of photodoping in conventional semiconductors (e.g. Si \cite{Meng2015}), or indeed larger-gap Mott/charge-transfer insulators \cite{iwai03,okamoto07}, where the ES response is dominated by an intraband Drude(-like) response of charges in the partially filled bands. %
While such mobile-carrier responses may deviate from a simple Drude conductivity (with a maximum of $\sigma_1$ at $\omega=0$), an inspection of the spectra in Fig.~\ref{fig:fluence}(b) at low fluence would require a strong suppression of the low-frequency response, and not only a moderate shift of the Drude peak away from $\omega=0$.
While various models have been proposed for non-Drude responses, including the effects of phonon scattering \cite{smith01,pustogow21}, generalized Fermi-Liquid behavior \cite{berthod13}, localization \cite{fratini14,fratini21}, including approaches explicitly accounting for electron correlations (e.g., based on DMFT theory) \cite{merino2000,merino2008,ferber14,fratini21,ahn22}, we are not aware of a model treatment (or experimental situation) to account for the spectra in Fig.~\ref{fig:fluence}(b) in terms of a mobile-carrier contribution.
Note that we tested thoroughly alternative spectral analyses, with an additional Drude(-Smith) conductivity component (as opposed to modifying the BE component), but these produced a significantly stronger \dRrel than observed at low frequencies, especially for low fluence.
Note that for our BE model, the finite \dRrel for $\omega\rightarrow 0$ is rather a result of the change in $\Delta\epsr(\omega\rightarrow 0)$ associated with the BE band.
Nevertheless, we cannot rule out a finite contribution to $\sigma'(\omega)$ from mobile doublons/holons, especially for large \Fex, where the spectral edge approaches $\omega=0$.
Note that one can estimate a DC photoconductivity $\sigma'(0)$ based on previous GS transport \cite{pinteric15} and Hall \cite{tanatar97} measurements.  At the highest temperature $T=100\Kel$ with available data along the $c$-axis, we have $\sigma(0)=2.4\pOhmcm$ for a Hall carrier density $\usr{N}{th}=4.5\cdot 10^{19}\pcmc$.  At our maximum predicted ES density, based on the inter-Hubbard bleach, 
$\Nex'=\tfrac{1}{2}\usr{N}{d}(1-\usr{W'}{H}/\usr{W}{H})=0.28\cdot\usr{N}{d}=1.7\cdot 10^{20}\pcmc$, one expects $\sigma'(0)=9.4\pOhmcm$.  A Drude contribution of this magnitude would be negligible in Fig.~\ref{fig:fluence}(b). %

Hence, in the absence of a theoretical model which would attribute such a low-frequency spectral response to the excited holon/doublons, the gap-closure model offers a compelling hypothesis, especially considering its reasonable quantitative account of our results, and previous assertions about Mott-gap closure upon photodoping \cite{golez15,yamakawa17,grandi21,murakami_arxiv23}. %
One notable issue concerns the observed conservation of spectral weight 
$\usr{W'}{H}+\usr{W'}{BE}\approx \usr{W}{H}$ vs. fluence.
Such a result would be quite intuitive for the case where the ES spectral weight is associated with the excited carriers, e.g. from a Drude response.  However, here we assert that the BE component is due to remaining singlons in the LHB, whose population is depleted by the excitation.
Even in the general Green function treatment of conductivity response based on a self-energy function \cite{merino2000,mutou2006,deng13} (typically obtained via DMFT), the common formulation of the intra-/interband dynamic conductivity (neglecting vertex corrections) would predict a reduction of the interband spectral weight due to depletion of the LHB, assuming the respective integrated density of states (DOS) of the LHB and UHB remain equal (as expected in the GS for the half-filled-band insulator case here).  This raises a compelling issue to inspect in  future theoretical treatments of the photodoped state in \kCl.
We finally note one last aspect of the results in Fig.~\ref{fig:fluence}: if one assumes the gap $\Delta$ indeed diminishes with increasing excitation density, one might expect that the initial multiplication of excitations $\Nex'/\Nex=\hbar\wex/\Delta$ could exhibit a non-linear growth with \Fex, although one sees rather a linear trend followed by saturation in Fig.~\ref{fig:fluence}(c) for $\usr{W'}{H}$.
Here, time-dependent non-equilibrium theory \cite{eckstein13,werner14} would be valuable to follow the temporal development of the excitation multiplication process in concert with gap closure.


\subsection{Relaxation dynamics and temperature dependence}

\begin{figure}[t]
\centering
\includegraphics
[width=0.85\linewidth]
{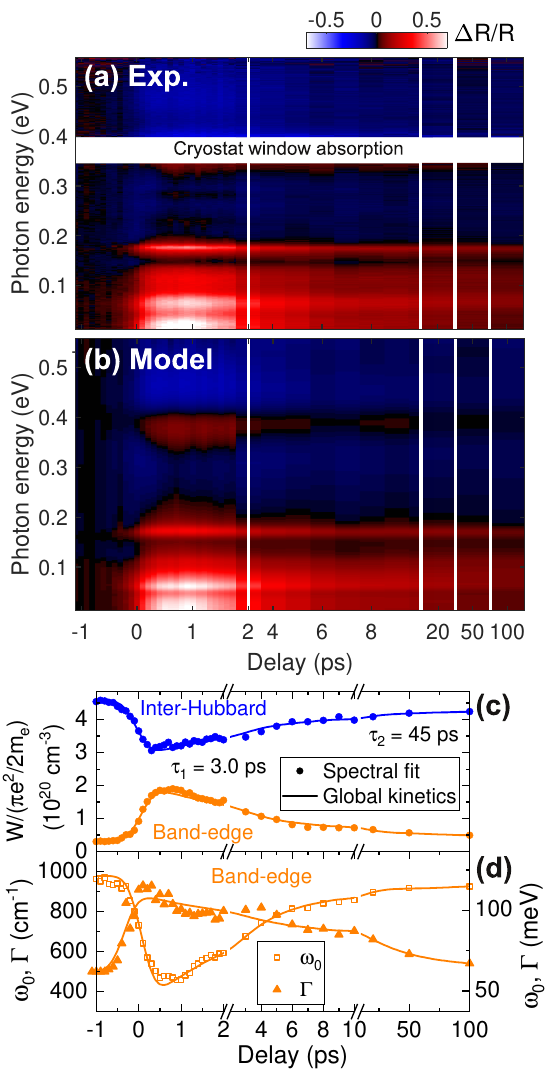}
\caption{
(a) Experimental and (b) fitted model excited-state differential reflectivity spectra for \kCl vs. excitation delay $\tau$ ($T=15\Kel$).
Selected model parameters (symbols) vs. $\tau$ from fitting in (b): (c) Spectral weights $W'(\tau)$ of inter-Hubbard and BE components; (d) BE frequency $\wz'(\tau)$ and damping $\Gamma'(\tau)$.  Included are bi-exponential global kinetic fits of each band parameter (curves), i.e. using $\tau_1=3.0\psec$ and $\tau_2=45\psec$ for all parameters (as well as a weak offset to account for long-lived thermal offsets).
} \label{fig:sgrams}
\end{figure}

In this last section, we investigate the time-dependence of the ES spectral response and relaxation dynamics following the initial sub-ps electronic processes.
In Fig.~\ref{fig:sgrams}(a), we plot a full 2D set of transient spectra $\dRrel(\omega,\tau)$ for $T=15\Kel$.
One can observe the same spectral features as shown in the spectral curves in Fig.s~\ref{fig:gs}(g) and \ref{fig:fluence}(a).
As per the last section, we apply our ES spectral analysis to each spectrum vs. $\tau$, to produce the 2D model spectra in Fig.~\ref{fig:sgrams}(b), which are seen again to reproduce the experimental data well.
Selected model band parameters vs. $\tau$ are shown in Fig.~\ref{fig:sgrams}(c,d).
In Fig.~\ref{fig:sgrams}(c), one observes the transient bleach of the inter-Hubbard band and concomitant growth of the BE component, while Fig.~\ref{fig:sgrams}(d) shows the gap-closure signature and broadening in terms of the transient BE peak/damping.
Note that for these measurements, the required \Fex to achieve a given magnitude of the Hubbard bleach was significantly higher than for the results in Fig.~\ref{fig:fluence}, which appears to be connected to  surface inhomogeneity.  Nevertheless, the behavior of, and relation between the fitted band parameters is very close to those in the last section. Moreover, having access to the absolute bleach of the inter-Hubbard band in each case allows to calibrate for these variations. Compared to other measurements, we arrive at a nominal fluence $\Fex\simx 180\muJcm$ for the data in Fig.~\ref{fig:sgrams}.%

In order to extract time-scales for the relaxation, we fit this set of parameters using a multi-exponential fit, e.g. 
$\usr{W'}{H}(\tau)=\usr{W}{H}+\Sigma_j\, \Delta W_{\mathrm{H}j}\cdot e^{-\tau/\tau_j}\Theta(\tau)$
(likewise for $\usr{W'}{BE}$, $\wzX{BE}'$ and $\usr{\Gamma'}{BE}$), convolved with a Gaussian time-response function $\propto e^{-2\tau^2/T_0^2}$ to account for the finite rise-time (here $T_0=370\fsec$).  We found that one can describe the kinetics of all four band parameters with the same two global time constants, i.e. a dominant fast relaxation with $\tau_1=3.0\psec$, and a weaker residual component with $\tau_2=45\psec$ (as well as small offsets to account for temperature changes which persist for much longer than the 100-ps time window). 
The fitted curves are also included in Fig.~\ref{fig:sgrams}(c,d).  The fact that all parameter kinetics vary with common time constants provides good support for our spectral analysis, and indicates that the evolution of the ES follows essentially a single relaxation channel (addressed again below).  We note that we do not resolve any spectral anomalies about $\tau=0$, e.g. from short-lived Drude/coherence-peak species \cite{okamoto10,golez15}. %
Measurements as per Fig.~\ref{fig:sgrams}, but with 70\% lower fluence (nominally $\Fex\simx 60\muJcm$) shows that $\tau_1$ reduces significantly to $1.8\psec$ (see Supplementary). 
We address this issue below in relation to the phonon-mediated gap recombination model.

\begin{figure}
\centering
\includegraphics
[width=0.7\linewidth]
{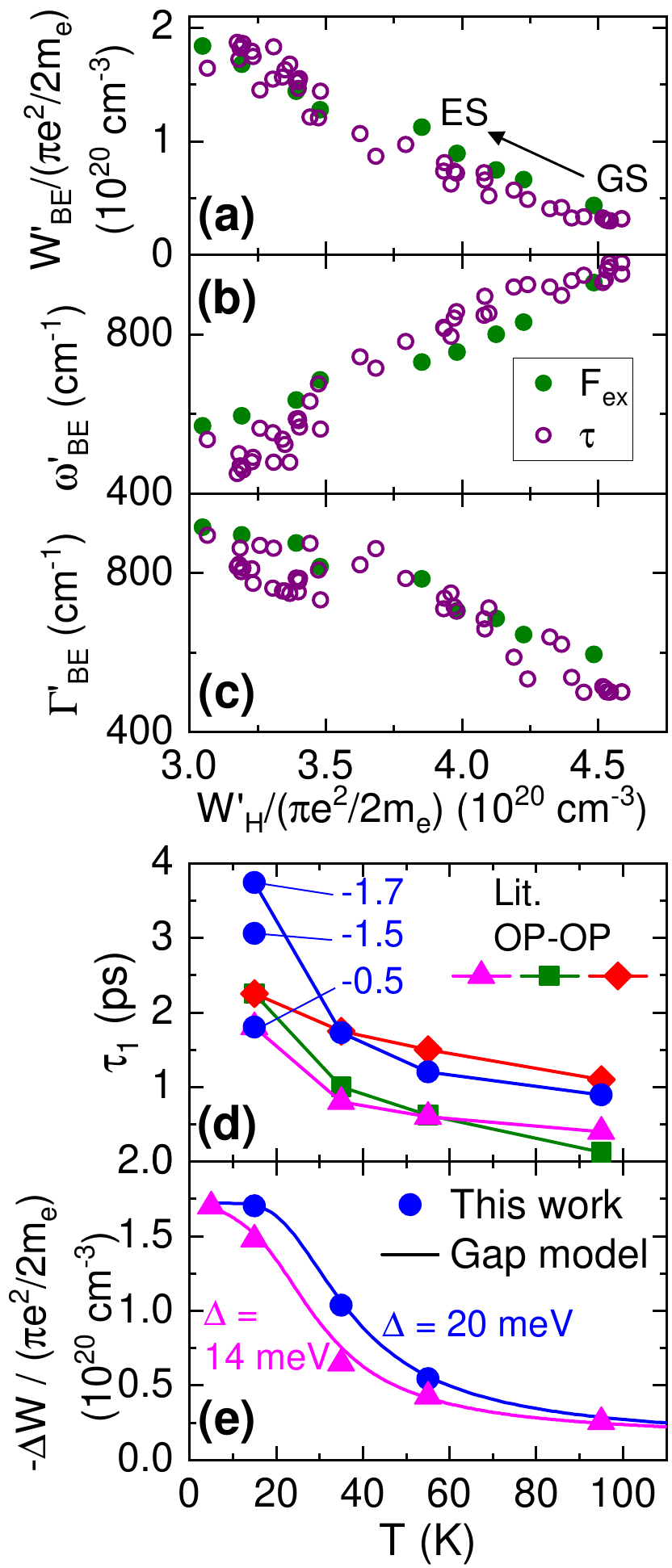}
\caption{
(a-c) Locus plots of ES BE parameters vs. corresponding inter-Hubbard band spectral weight for (i) initial ES vs. \Fex ($\tau=1\psec$) and (ii) during relaxation vs. $\tau$ for fixed \Fex: (a) spectral weight, (b) Lorentzian peak frequency, and (c) damping. 
(d,e) Temperature dependence of ES response parameters, including comparison to 
previous literature reports with all-optical pump-probe (triangles \cite{toda11}; squares \cite{naito08}; diamonds \cite{tsuchiya19}).
(d) Primary relaxation time constant $\tau_1$. For present study, three values are plotted for $T=15\Kel$ from measurements with different excitation density, and are labeled with their respective initial inter-Hubbard spectral weight bleach values $\Delta\usr{W}{H}/(\pi e^2/2\usr{m}{e})$ (in units of $10^{20}\pcmc$). %
(e) Photoinduced population: for present THz-IR spectral analysis using bleach inter-Hubbard spectral weight; for literature OP-OP using relative amplitudes of \dRrel signal.
Also included are model curves based on gap recombination model  \cite{kabanov99,demsar99,kabanov05}, with $\Delta=20\meVolt$ (this work) and $\Delta=14\meVolt$ \cite{toda11}. 
} \label{fig:FtauTdep}
\end{figure}

As mentioned in the Introduction, for such strongly correlated electronic systems, the degree of photodoping may well distort the energy-state distribution such that it affects the relaxation pathways, especially when coupling to phonons (and spins) also plays a role.  To test for such effects, in Fig.~\ref{fig:FtauTdep}(a-c), we compare the inter-dependence of the model BE parameters vs. the inter-Hubbard bleach, for both the initial state vs fluence (at $\tau=1\psec$, as per Fig.~\ref{fig:fluence}), and as a function of delay $\tau$ (Fig.~\ref{fig:sgrams}(c,d)).  (Note that the data are from the same measurement run to ensure consistency).
As can be seen, the loci for all three BE parameters are essentially superposed between the two situations.  Hence, during relaxation, the system progresses through states very similar to the initial states at the corresponding excitation densities.  This shows that although the energy-state distribution may be strongly distorted at our  photodoping concentrations, this is effectively ``retraceable'' during relaxation, as opposed to causing the system to take qualitatively different relaxation pathways vs. initial photodoping.

We finally investigate the GS $T$-dependence of the photodoping response. %
Previous studies employing all-optical pump-probe \cite{naito08,toda11,tsuchiya19} (using pulses at single wavelengths to probe the ES density via Pauli-blocking of higher-energy transitions into the UHB) showed a clear $T$-dependence for both the initial population \Nexm (at fixed \Fex after the sub-ps processes) and the subsequent relaxation time scale $\tau_1$.
While the ES electronic temperature is predicted to be on the many-100-K scale (after electron thermalization), one can expect this $T$-dependence due to considerations of the partition of the excitation energy into electrons (residing directly above the gap $\Delta$) and phonons, relative to their respective thermal (GS) populations before excitation.
This is described by the phonon-mediated gap recombination  model \cite{demsar99,kabanov99,kabanov05} which has had much success quantitatively accounting for the  dependence of $\Nexm(T,\Fex)$ on $\Delta(T)$ (as well as the subsequent phonon-mediated recombination) for the excited state in gapped systems. %
The model is based on the statistics of the ES carriers and high-energy phonons in quasi-equilibrium (before subsequent irreversible quasi-particle recombination, which is controlled by the decay of high-energy phonons to lower energy ones, i.e. the phonon bottleneck). Here one considers only phonons with energy $\hbar\omega_q>\Delta$ ($\Delta$ defined here as the full gap for a single doublon-holon excitation/recombination event, as opposed to $2\Delta$ as typical, e.g. for Cooper pairs in superconductors \cite{demsar99}). 
While one can derive the quasi-equilibrium densities \Nexm and $\usr{N'}{ph}$ from integrating over the statistical energy distributions for a given ES temperature $T'$, the scaling can be obtained directly from detailed balance \cite{lawrence99}, i.e. $\usr{N'}{ph} \propto (\Nexm)^2$, which results from the fact that the scattering rate obeys
$\partial_t \usr{N'}{ph} \propto \usr{N'}{doub}\usr{N'}{hol}=(\Nexm)^2$.  Upon excitation, one then has 
$\delta\usr{N'}{ph} \propto 2\Nexm\cdot \delta\Nexm$ and the ratio $\delta\usr{N'}{ph}/\delta\Nexm$ grows with the thermally excited density \Nexm, such that more excitation energy is partitioned to the phonon bath with increasing GS temperature $T$ and $\delta\Nexm$ is reduced, according to 
$\delta\Nexm=(\hbar\wex/\Delta)/(1+g(\Delta,T)e^{-\Delta/2kT})$ \cite{kabanov99} where $g(\Delta,T)$ is assumed to be a slowly varying function of $T$ dictated by the electronic and phonon DOS. %
(Note that in previous sections, we did not distinguish between $\Nex'$ and $\delta\Nex'$, as they are essentially equal at low $T$).

Fig.~\ref{fig:FtauTdep}(d,e) show the $T$-dependence of the (d) primary relaxation time constant $\tau_1(T)$ and (e) initial Hubbard bleach $\usr{\Delta W}{H}(\tau=1\psec)$, along with data from previous reports using single-wavelength all-optical pump-probe (OP-OP).
As mentioned above, we find that $\tau_1$ depends also on \Fex, and for $T=15\Kel$ we include three values of the fitted $\tau_1$, for different initial $\usr{\Delta W}{H}$ in Fig.~\ref{fig:FtauTdep}(d).
One sees that all results follow approximately a $\tau_1\propto 1/T$ dependence, which is predicted from the phonon-mediated gap model \cite{kabanov99}, and similar to relaxation in metals can be traced to the rate at which high-energy phonons decay to low-energy ($\hbar\omega_q <\Delta$) ones (phonon-bottleneck).
While in \cite{toda11}, it was found that $\tau_1$ was approximately constant vs. $\Fex\rightarrow 400\muJcm$, here we already see a slowing down at fluences below $200\muJcm$. 
This could be due to the different excitation photon energy $\hbar\wex=0.95\eVolt$ \cite{toda11}, but could also be due 
to the fact that we directly probe the LHB/UHB species via their low-energy excitations (as opposed to indirectly via higher-energy interband bleach in OP-OP).  
While a longer relaxation time at higher fluence could be due to different physical effects (but not due to a bimolecular holon-doublon recombination rate, which would have the opposite trend), it is indeed predicted qualitatively by the gap recombination model, if one considers that the gap $\Delta$ decreases with increasing \Nexm.  For a smaller gap energy $\Delta$, the DOS for phonons with energy above (below) $\Delta$ increases (decreases), i.e. the rate for high-energy phonons to decay into low-energy ones will decrease, which also determines the decay rate for \Nexm in this model.

Our main test of the gap model vs $T$ concerns the initial excited-state density $\delta\Nex'$, as depicted by $\usr{\Delta W}{H}$ in Fig.~\ref{fig:FtauTdep}(e), which is seen to decay with increasing $T$.  Fitting the equation above to our data (for $T \leq 55\Kel$) yields the model curve shown, using a value $\Delta=20\meVolt$, in reasonable agreement with the gap value from DC transport.
We also include comparative results from \cite{toda11} (relative initial OP-OP \dRrel signal amplitude), which are similar but show a somewhat faster decay with $T$ (with a correspondingly smaller fitted value of $\Delta=14\meVolt$).
(Note that the numerical value of $\Delta=7\meVolt$ given in \cite{toda11,tsuchiya19} is based on the definition of the excitation gap as $2\Delta$).
Again, this may result from the different excitation/probing in the two cases.  Nevertheless, our estimate seems to be closer to other independent estimates of $\Delta$.

\section{Conclusions}

We have proposed a distinct scenario for the response of the Mott insulator \kCl to ultrafast photodoping, based on a rigorous analysis of transient reflectivity spectra covering the full THz and mid-IR ranges.
The results put forward a quantified measure of progressive gap closure vs. excitation density, which can be used to test ongoing theoretical descriptions of both the GS and photodoped ES in such narrow-gap Mott systems.
Here, further efforts to account for the ground-state DC and dynamic conductivity vs. $T$ would be already be an important step, while treatments of the non-equilibrium response may shed more light on the coupling of the correlated electron system to phonon and spin degrees of freedom, as well as the possible role of other effects such as disorder and frustration.

\section{Acknowledgments}
We gratefully acknowledge funding by the German Research Foundation (DFG) via the Collaborative Research Center TRR 288 (422213477, projects A06 and B08).
We thank Jure Demsar, Vladimir Dobrosavljevic and Philipp Werner for helpful discussions.


%

\end{document}


\title{Supplementary material to: \\ \thetitle}
\author{Konstantin Warawa}
\affiliation{\ffm}
\author{Yassine Agarmani}
\affiliation{\ffm}
\author{Harald Schubert}
\affiliation{\ffm}
\author{Martin Dressel}
\affiliation{\stutt}
\author{Michael Lang}
\affiliation{\ffm}
\author{Hartmut G. Roskos}
\affiliation{\ffm}
\author{Mark D. Thomson}
\affiliation{\ffm}

\maketitle

\section{Experimental setup} \label{sec:exp}

\begin{figure*}[!t]
\centering
\includegraphics
[width=0.55\textwidth]
{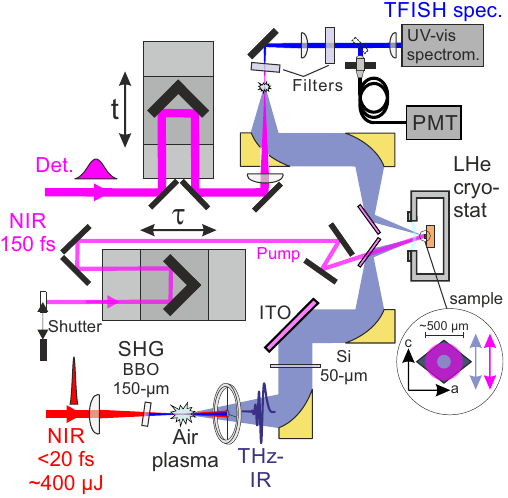}
\caption{Schematic of experimental optical-pump Terahertz-infrared (THz-IR) probe setup, including THz-IR generation, sample focus and detection regions. See text for details and abbreviations.
} \label{fig:setup}
\end{figure*}

A schematic of the setup for the optical-pump THz-infrared-probe experiments is shown in Fig.~\ref{fig:setup}, based on 
a 1-kHz \tisa amplifier laser (Clark-MXR CPA-2101, $\lambda_0=775$~nm, pulse duration $\simx 150\fsec$, pulse energy $900\muJ$). As described in e.g. \cite{thomson10, thomson23}, the near-IR (NIR) pulses are spectrally broadened by propagation through an Ar-gas-filled capillary (2.1 bar) and subsequently temporally compressed using negative-dispersion mirrors, leading to (near)-transform-limited sub-20-fs pulses with $\simx 400\muJ$ pulse energy. These are used to generate the ultrabroadband THz-IR probe pulses (with bandwidth coverage from $\simx 0$-$0.6\eVolt$) via the two-color air plasma emission technique, where a Type-I $\beta$-BBO (150\mum thickness) is used for second-harmonic-generation (SHG). An axial stop is placed directly after the air plasma to block the plasma pump light, while the probe pulses are re-collimated and focused, both onto sample and subsequent detection regions, by off-axis paraboloidal mirrors. Any residual plasma pump light is suppressed by a 50-µm thick Si wafer before focusing on the sample.
An indium-tin-oxide (ITO) beam-combiner is used to allow a collinear beam geometry for 150-fs NIR pump and THz-IR probe pulses on the sample  (with relative delay $\tau$, and incidence angle $\theta=18^{\circ}$). The pump beam size is sufficiently large to achieve a homogeneous excitation of the sample surface over the probe beam profile.

In the detection region, the THz-IR pulses reflected from the sample co-propagate with 150-fs NIR gate pulses (with relative delay $t$), enabling the measurement of terahertz-field-induced second-harmonic (TFISH) spectrograms $I(\omega,t)$ \cite{thomson23}. A half-wave plate and polarization beamsplitter are utilized for simultaneous measurements of spectra (with an UV-vis spectrometer) and spectrally integrated TFISH signals (with a photomultiplier tube (PMT)), the latter allowing for fast optimization and tracking of the TFISH reference signal during measurements. Due to both minimal dispersion in the THz-IR beam path and the relatively long NIR detection pulse duration, the full probe spectral range (with frequency resolution of $\simx 10$~meV) can be measured at a single constant delay $t=t_0$.

The sample is attached to a copper disk with a single small drop of Apiezon N adhesive in one corner (avoiding potential strain effects upon cooling) and mounted in the liquid-helium (LHe)-flow cryostat. The cooling rate in these experiments is $\simx 1.5\Kel/\mathrm{min}$, which in the $T$-range between $65$-$80\Kel$ near the glass-like transition is reduced to $\simx 0.1\Kel/\mathrm{min}$ to minimize any structural disorder in the Mott-insulating phase of \kCl \cite{hartmann14}.

\newpage

\section{Expression for the complex field reflectivity with longitudinal excitation profile}

As detailed in the main paper, to calculate the model reflectivity $r'(\omega)$ of the excited sample surface (for a given complex conductivity $\sigma'(\omega)$), we account for the finite excitation depth $\usr{\delta}{ex}=1/\alphaex$, which becomes important for longer wavelengths in the THz range.

One assumes the excitation leads to a depth profile for the relative permittivity $\epsr'(z,\omega)$ given by
\begin{equation}
\epsr'(\omega,z)=\epsr(\omega) + 
\Delta\epsr(\omega)\cdot e^{-\alphaex z}
\label{eq:epsprofile}
\end{equation}
where $\epsr(\omega)=\epsri-i\sigma/\varepsilon_0\omega$  
(likewise for $\epsr'$, $\sigma'$) and $\Delta\epsr=\epsr'-\epsr$.

The expression $r'$ for normal incidence is well-known in the literature, and has been employed in e.g. \cite{Meng2015,thomson17}. 
For s-polarization at oblique incidence angle $\theta$, the wave equation for the THz-IR field $E(\omega,z)$  becomes \cite{khachatrian15}:
$$\partial_z^2E + k_0^2(\epsr'-\sin^2{\theta})E=0,$$
or, substituting Eq.~\eqref{eq:epsprofile}:
$$\partial_z^2E + (a^2-b^2e^{-\alphaex z})E=0, 
\qquad a^2=k_0^2(\epsr-\sin^2{\theta}), 
\quad b^2=-k_0^2\Delta\epsr.$$
As per the case $\theta=0$, this can be transformed to the Bessel differential equation via substitution of $Z=2b\usr{\delta}{ex}e^{-z/2\usr{\delta}{ex}}$, to yield a solution of the same form as for $\theta=0$ (only with the new definition of $a$), i.e.
\begin{equation}
    r'=\frac{1-\usr{n}{eff}}{1+\usr{n}{eff}}, \qquad
    \usr{n}{eff}=n+\sqrt{\Delta\epsr}\cdot X_\beta
\end{equation}
where $n=\sqrt{\epsr}$, 
$X_\beta(\xi)=I_{\beta+1}(\xi)/I_{\beta}(\xi)$ is the ratio of modified Bessel functions with $\beta=2i\omega n(c\alphaex)^{-1}$ and 
$\xi=2i\omega\sqrt{\Delta\epsr}(c\alphaex)^{-1}$.

\newpage

\section{Temperature-dependent ground-state spectra of \kCl} \label{sec:gstemp}


\begin{figure*}[!htb]
\centering
\includegraphics
[width=0.87\textwidth]
{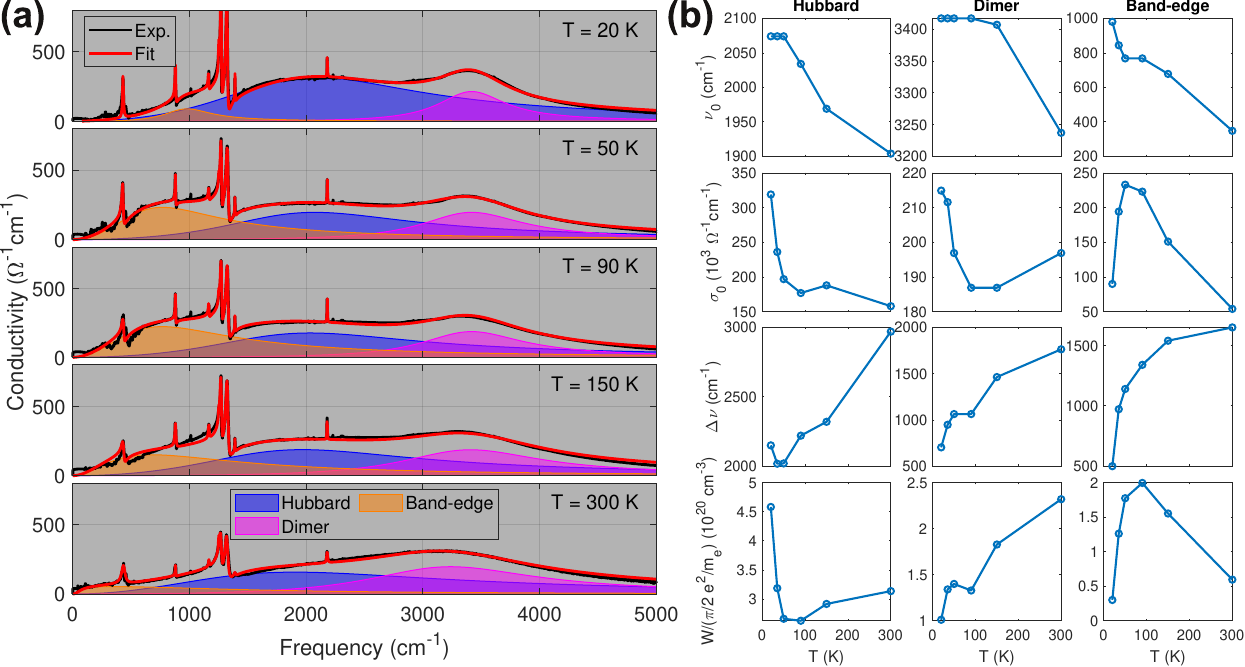}
\caption{(a) Experimental ground-state conductivity spectra of \kCl (black, originally published in \cite{faltermeier07}) and multi-band Lorentzian fit (red) for different temperatures $T$ between $20\mbox{-}300\Kel$, with polarization parallel to c-axis. (b) Corresponding parameters of the inter-Hubbard, intra-dimer and band-edge components vs. $T$, including center frequencies $\nu_0$, band conductivities $\sigma_0$, bandwidths $\Delta\nu$ and spectral weights W (normalized to equivalent density assuming free electron mass).
} \label{fig:gsnodrude}
\end{figure*}

\begin{figure*}[!htb]
\centering
\includegraphics
[width=0.87\textwidth]
{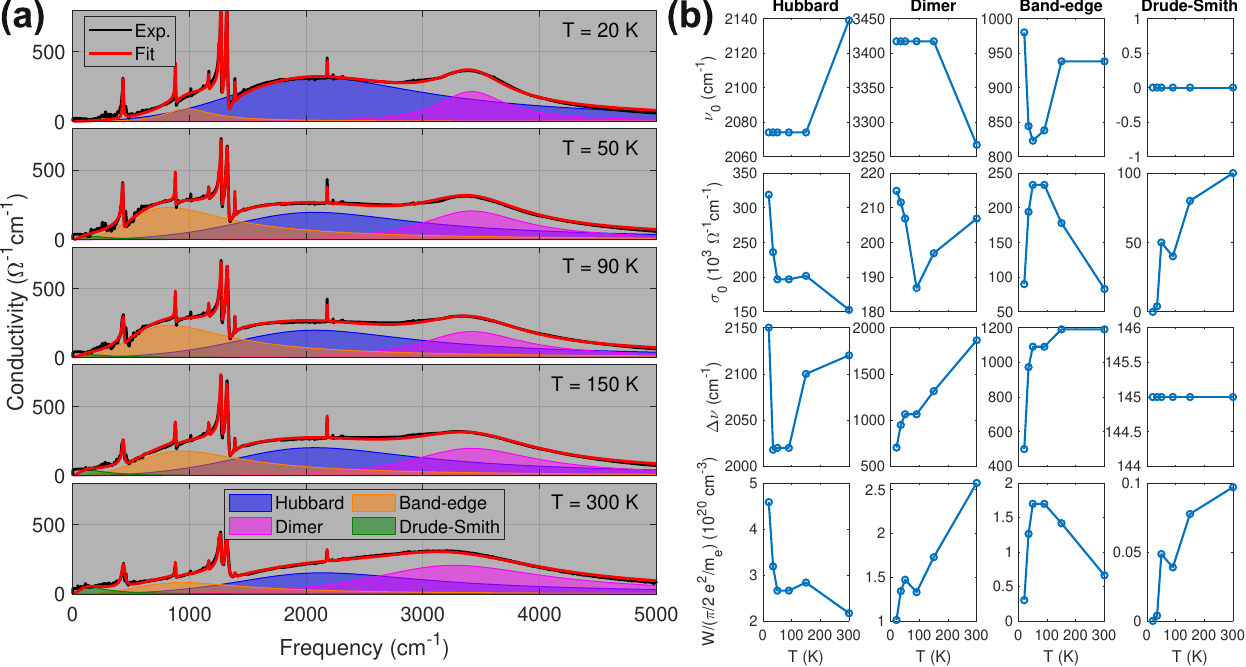}
\caption{$T$-dependent spectra and parameters as per Fig.~\ref{fig:gsnodrude}, but including an additional Drude-Smith contribution (with backscattering parameter $c_{\mathrm{DS}}$ \cite{smith01}). The set band conductivity values $\sigma_0$ are based on DC conductivities $\sigma_{\mathrm{DC}}$ obtained by transport measurements \cite{pinteric15}, where $\sigma_{\mathrm{DC}} = \sigma_0 \cdot (1+c_{\mathrm{DS}})$, $c_{\mathrm{DS}}=-0.99$ below $T = 90\Kel$ and $c_{\mathrm{DS}}=-0.95$ at/above $T = 90\Kel$.
} \label{fig:gswithdrude}
\end{figure*}

\newpage

\begin{figure*}[ht]
\centering
\includegraphics
[width=0.7\textwidth]
{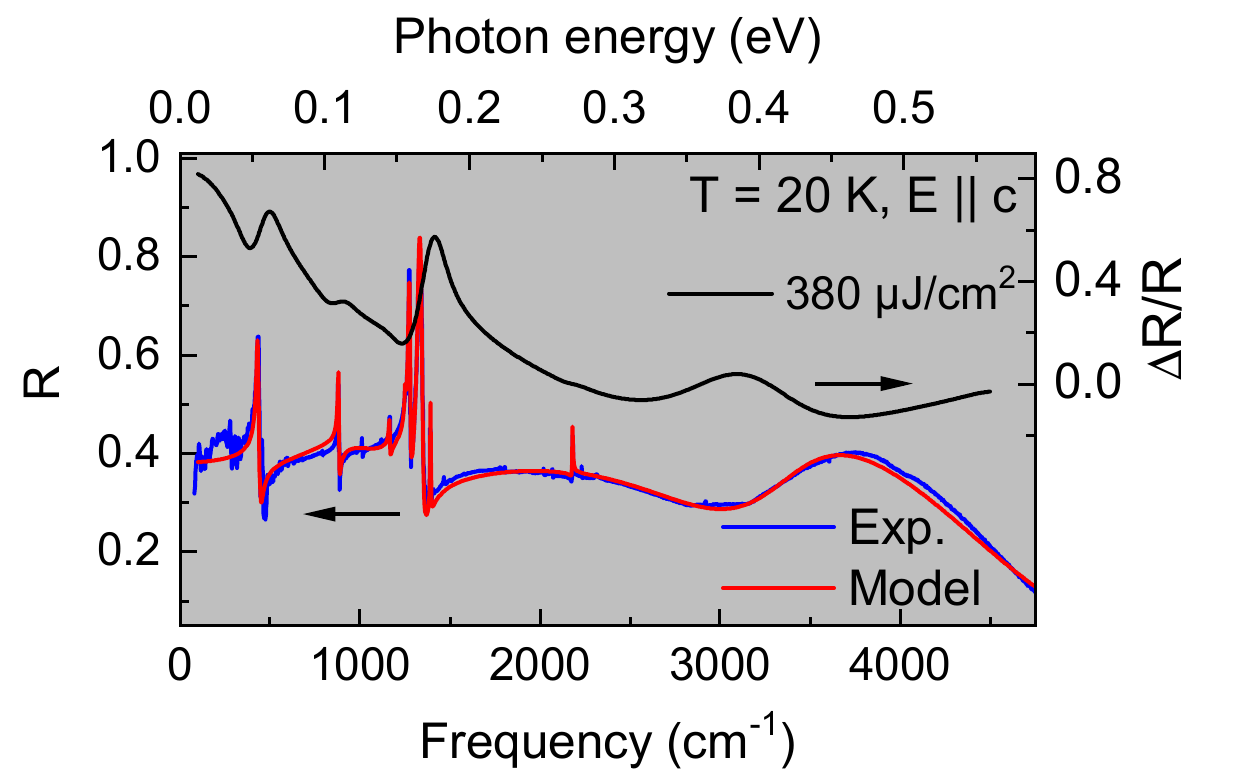}
\caption{Experimental ground-state reflectivity spectrum of \kCl at $T=20\Kel$ (blue, originally published in \cite{faltermeier07}) and multi-band Lorentzian fit (red), with polarization parallel to c-axis and corresponding to the conductivity spectrum in Fig.~\ref{fig:gsnodrude}(a). An exemplary excited-state differential reflectivity spectrum \dRr at $T=15\Kel$, $\Fex \simx 380\muJcm$ and delay $\tau=1\psec$ after $1.6\eVolt$-excitation (Fig.~1(g) in main paper) is shown additionally.
} \label{fig:gsrefl}
\end{figure*}

 \newpage
\section{Analysis of Excited-state differential reflectivity spectra on the basis of a Drude-Smith band (without gap closure)}

\begin{figure*}[ht]
\centering
\includegraphics[width=0.9\textwidth]
{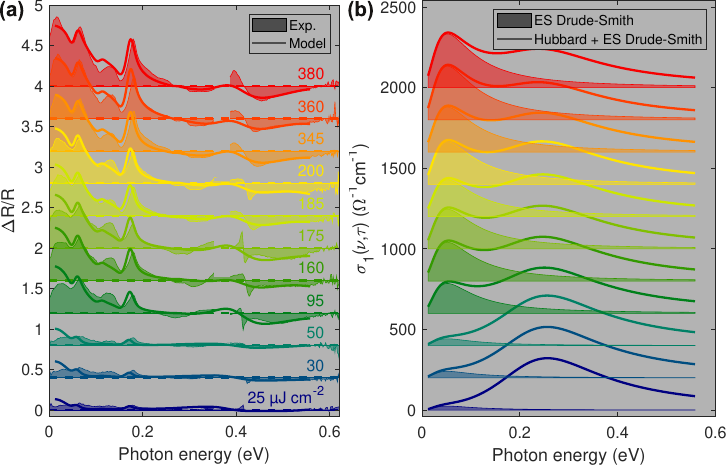}
\caption{
Excited-state spectra vs. excitation fluence $\Fex$ for $T=15\Kel$ and delay $\tau=1\psec$ after excitation, as per Fig.~2(a,b) in main paper, only using Drude-Smith to model low-frequency photoinduced spectral weight.
Drude-Smith model parameters as per Fig.~\ref{fig:gswithdrude} above. 
(a) Differential reflectivity spectra $\dRr$ (\Fex labeled for each curve, vertically staggered for clarity), and model fits. The low-frequency deviation for \dRrel is clearly evident for intermediate fluence values. %
(b) Corresponding model conductivity spectra: Lorentzian components from Drude-Smith alone (filled regions) and sum of Drude-Smith and band-edge components (curves). For the highest fluence values, the band-edge model (Fig.~2 in main paper) converges to a very similar conductivity function.
} \label{fig:esfluenceDS}
\end{figure*}

\newpage
\section{Transient 2D spectra for lower excitation fluence}

\begin{figure*}[!h]
\centering
\includegraphics[width=0.55\textwidth]
{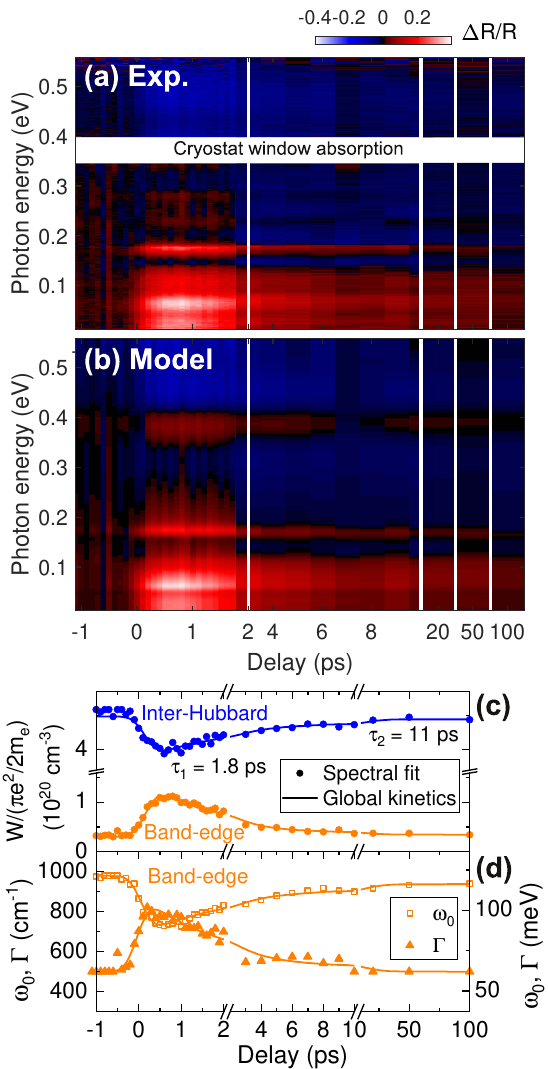}
\caption{
(a) Experimental and (b) fitted model excited-state differential reflectivity spectra vs. excitation delay $\tau$ ($T=15\Kel$), with lower fluence (compare with Fig.~3 in main paper).
Selected model parameters (symbols) vs. $\tau$ from fitting in (b): (c) Spectral weights $W'(\tau)$ of inter-Hubbard and band-edge components; (d) band-edge frequency $\wz'(\tau)$ and damping $\Gamma'(\tau)$.  Included are bi-exponential global kinetic fits of each band parameter (curves), i.e. using $\tau_1=1.8\psec$ and $\tau_2=11\psec$ for all parameters (as well as a weak offset to account for long-lived thermal offsets).
} \label{fig:essgramlowF}
\end{figure*}

\newpage
%